\documentstyle[prl,aps,epsfig,twocolumn]{revtex}

\begin{document}
\draft

\twocolumn[\hsize\textwidth\columnwidth\hsize\csname@twocolumnfalse\endcsname
\title{Engineering Progressive Decoherence with Quantum Jumps  in Charge Qubit}
\author{Y. D. Wang, Y. B. Gao, C. P. Sun$^a$}
\address{Institute of Theoretical Physics,the Chinese Academy of Science,
 Beijing, 100080, China}
\maketitle

\begin{abstract}
For the Josephson junction charge qubits  with macroscopically
quantum natures, we propose a theoretical scheme to observe the
loss of quantum coherence  through coupling such qubit system to
an engineered reservoir, the harmonic oscillator mode in the LC
circuit formed by the inductor and the separated capacitors.
Similar to  the usual cavity QED system in form, this charge qubit
system with engineered couplings  shows  the quantum jumps
(C.P.Sun et al Fortschr. Phys. \textbf{43}, 585 (1995) )in a
progressive decoherence process. Corresponding to two components
of superposition of two charge states, the inductor evolves
simultaneously towards two distinct quasi-classical states
entangling with two states of the charge qubit. Then it induces
the quantum decoherence for the induced squeezing macroscopically
in the LC mode.
\end{abstract}
]

It is well-known that the superposition of quantum states lies at the very
heart of modern quantum theory. In an ideal situation the quantum coherence
implied by this superposition results in various dramatic features in
quantum mechanics \cite{wheeler-zurek}. However, the real systems are never
isolated completely from the surrounding environment. The interaction with
the environment (a reservoir ) or other external systems will lead to the
entanglement between them, and then the randomness or the classicality of
environment will wash out the phases of quantum system \cite{caldeira-legget}%
. This consideration explains why the quantum superposition does not seem to
appear in the macroscopic world: there happens the transition from the
quantum world to classical world \cite{zurek1}.

This issue is directly related to quantum measurement problem where the
coupling of the measured system with the measuring apparatus ( detector)
will cause the reduction of superposition or wave packet collapse \cite{sun1}%
. It should be emphasized that the coupling between the measured system and
the detector can be controlled to satisfy one's need in measurement. This is
quite different from the coupling with the environment, the detailed
knowledge of which is usually inavailable. Actually in the past few years,
the cavity QED system \cite{brune} and the laser cooled trapped ions \cite%
{maytt} were utilized to demonstrate how to "engineer" the system-reservoir
coupling so that the progressive decoherence can be observed with
experimentally accessible technologies. In this letter, we show that, in the
"qubit way"--a two level approximation \cite{shnirman,nakamura,mooij}, a
solid system --the Josephson junction can also implement the engineered
system-reservoir interaction to illustrate the detailed dynamics of quantum
decoherence. In fact, in the most recent experiments of charge and flux
qubit of Josephson junction, the much longer time Rabi oscillation with very
large qubit quality factor $Q_{\phi }=\tau _{\phi }\omega $( $2.5\times
10^{4}$ for charge qubit \cite{vion} and $2\times 10^{15}$for flux qubit 
\cite{yu}, $\tau _{\phi }$ is the decoherence time and $\omega $ is the
"Larmor precession frequency") is observed. These physical realizations of
qubit offer us the possibility of manipulating the quantum states of the
mesoscopic electrical circuit and engineering the coupling between the qubit
and the artificial environment. Most recently, the relaxation and dephasing
that result from the control and the measurement setup itself in experiments
have been discussed for the Josephson persistent-current qubits\cite{mooij1}%
. In this letter we will pay our attention to the charge qubit.

\begin{figure}[tbp]
\caption{A charge qubit of tunable coupling is connected with an inductor to 
$L$, and gate voltage $V_{g}$ can be controlled to adjust the coupling of
the Cooper pair with its engineered reservoir. }
\end{figure}

For implementing the engineered reservoir couplings, one choice is to
connect the Josephson junction charge qubit to a LC-oscillator formed by
adding an inductor with tunable inductance $L$.(see Fig. 1). Here, the
charge qubit of tunable coupling is a complex Cooper pair box formed by a dc
SQUID with two symmetric junction. $C_{J}$ is the capacitor of tunnel
junction, $E_{J}$ the Josephson coupling energy, $C_{g}$ the gate capacitor
and $V_{g}$ the control gate voltage . The Hamiltonian of the total system
can be written down according to ref. \cite{maklin} as 
\begin{eqnarray}
H &=&\frac{q^{2}}{2C}+\frac{\phi ^{2}}{2L}+ \\
&&4E_{c}(n-n_{g})^{2}-E_{J}(\phi _{x})\cos (\theta -\eta ^{\prime }\phi )
\end{eqnarray}%
where $n$ is the number operator of excess Cooper-pair charges on the
island, and $\theta $ the phase of the superconductor order parameter, $\phi 
$ the flux through the inductor, $\phi _{x}$ the external flux and $q$ the
total charge accumulated on the gate capacitor. The others parameters are
defined as $C=\frac{C_{J}C_{g}}{C_{J}+C_{g}}$, $\eta ^{\prime }=\frac{2\pi }{%
\phi _{0}}\frac{C}{C_{J}}$, $E_{C}=\frac{e^{2}}{2(C_{J}+C_{g})}$,$n_{g}=%
\frac{C_{g}V_{g}}{2e}$ $E_{J}(\phi _{x})=2E_{J}^{0}\cos (\pi \frac{\phi _{x}%
}{\phi _{0}})$, and $\phi _{0}=\frac{h}{2e}$ denotes the flux quanta.

To form a qubit or a two-level system, one need to tune the gate voltage $%
V_{g}$ so that $n_{g}$ is approximately a half-integer. In this case the
charge eigen-states $|0\rangle _{c}$ and $|1\rangle _{c}$ are approximately
degenerate and the other energy levels are far from these two states. In the
case of weak coupling $\frac{C}{C_J}\sqrt{\langle\phi^2\rangle}\ll 1$, one
can keep $\phi$ to the first order and "isolate" $|0\rangle _{c}$ and $%
|1\rangle _{c}$ to implement a qubit system with Hamiltonian \cite{maklin} 
\begin{equation}
H=\hbar \omega a^{\dag }a-\frac{1}{2}\hbar \omega _{a}\sigma
_{z}+i(a-a^{\dag })\hbar g\sigma _{y}
\end{equation}%
with three crucial parameters $\omega =\sqrt{\frac{1}{CL}}$ , $\omega _{a}=%
\frac{1}{\hbar }\sqrt{16E_{c}^{2}(1-2n_{g})^{2}+E_{J}^{2}}$ ,$g=\frac{\pi
E_{J}}{\phi _{0}\hbar }\frac{C}{C_{J}}(\frac{\hbar ^{2}L}{4C})^{1/4}$. Here,
we have introduced the creation and annihilation operators, $a^{\dag }$ $=%
\frac{1}{2}(\frac{4L}{\hbar ^{2}C})^{1/4}(q+i\sqrt{\frac{c}{L}}\phi $ $)\ $%
and $a.$The quasi-spin operators $\sigma _{z}=|0\rangle \langle 0|-|1\rangle
\langle 1|,\sigma _{y}=-i(|1\rangle \langle 0|-|0\rangle \langle 1|)$ and $%
\sigma _{x}=|1\rangle \langle 0|+|0\rangle \langle 1|$ are defined in the
rotation representation with the bases $|0\rangle =\cos \frac{\theta }{2}%
|0\rangle _{c}+\sin \frac{\theta }{2}|1\rangle _{c}$ and $|1\rangle =-\sin 
\frac{\theta }{2}|0\rangle _{c}+\cos \frac{\theta }{2}|1\rangle _{c}$ \ for $%
\tan \theta =E_{J}/[4E_{c}(1-2n_{g})]$. It is noticed that $|0\rangle
_{c}(|1\rangle _{c})$ physically represents the state of no (one) excess
cooper pair on the island.

The above model is quite similar to a cavity QED model without the
rotation-wave-approximation (RWA), \ which usually describes the single mode
cavity interaction with an off-resonance two-level atom \cite{Zhu}. In this
cavity QED model, when the detuning between the cavity frequency and the $%
|0\rangle \leftrightarrow |1\rangle $ transition frequency is large enough
to avoid any energy transfer between the atom and the cavity , the atoms in
different states $|1\rangle $ and $|0\rangle $ will modify the phase of
cavity field in different ways \cite{brune,sun2} and thus induce the quantum
decoherence of atomic states superposition. We can consider these issues
about decoherence in the present charge qubit system. The large detuning
condition $\gamma =\frac{g}{|\omega _{a}-\omega |}\ll 1$ is easily satisfied
by taking proper parameters in experiments \cite{shnirman,nakamura,maklin}.
For example, we can take $C_{J}\simeq 10^{-16}F,$ $C_{g}\simeq
10^{-16}F,L\simeq 5\times 10^{-6}H,E_{J}\simeq 0.05K$ . In this case we
estimate $\omega _{a}\simeq 8.06\times 10^{10}Hz,$ $\omega \simeq 4.47\times
10^{10}Hz,$ $g\simeq 2.57\times 10^{9}Hz.$ Then we have $\gamma =7\times
10^{-2}\ll 1$, and we need not invoke the rotation wave approximation.

With the above consideration for the rational parameters in the experiment,
we shall adiabatically eliminate coherence effect between $|0\rangle $ and $%
|1\rangle $. Then we obtain an effective Hamiltonian $H_{eff}=H_{1}|1\rangle
\langle 1|+H_{0}|0\rangle \langle 0|,$which is diagonal with respect to $%
|0\rangle $ and $|1\rangle $, and the effective actions on the LC circuit
from two qubit states $|0\rangle $ and $|1\rangle $ are 
\begin{equation}
H_{k}=(\omega +\frac{2g^{2}}{\Delta })a^{\dag }a+(-1)^{k}\frac{g^{2}}{\Delta 
}(a^{2}-a^{\dag }{}^{2})+\varepsilon
\end{equation}%
for $k=0,1$ respectively. Here, $\Delta =\omega _{a}-\omega ,\varepsilon =%
\frac{g^{2}}{\Delta }-(-1)^{k}\frac{\omega _{a}}{2}$. It is easy to see that 
$H_{eff}$ is a typical dynamics Hamiltonian creating entanglement of the
subsystems. In fact, starting from a factorized initial state $|\psi
(0)\rangle =(c_{0}|0\rangle +c_{1}|1\rangle )\otimes |s(0)\rangle $, the
total system driven by $H_{eff}$ will evolve into an entanglement state 
\begin{equation}
|\psi (t)\rangle =c_{0}|0\rangle \otimes |s_{0}(t)\rangle +c_{1}|1\rangle
\otimes |s_{1}(t)\rangle
\end{equation}%
where $|s_{k}(t)\rangle =$ $\exp (-iH_{k}t)|s(0)\rangle $ $(k=0,1)$ and $%
|s(0)\rangle $ is the initial state of the LC circuit. Therefore, a charge
state superposition in terms of $|0\rangle $ and $|1\rangle $ will cause the
LC circuit state to evolve along the two directions $|s_{1}(t)\rangle $ and $%
|s_{0}(t)\rangle $. The time evolution dominated by the conditional dynamics
Hamiltonian $H_{eff}$ means to implement an ideal pre-quantum measurement
when the overlapping $\langle s_{1}(t)|s_{0}(t)\rangle $ approaches zero 
\cite{zhang}. Physically the pre-measurement implies a quantum decoherence
of the subsystem formed by charge qubit. We consider the reduced density
matrix of the charge qubit at time $t$. Its off-diagonal elements are
determined by $c_{1}c_{0}^{\ast }\langle s_{1}(t)|s_{0}(t)\rangle $ and
vanishes completely as the overlapping $\langle s_{1}(t)|s_{0}(t)\rangle $
is zero. In this sense, the decoherence factor defined by $D(t)=|\langle
s_{1}(t)|s_{0}(t)\rangle |$ characterizes the extent of decoherence and the
time-dependent behavior of $D(t)$ means a progressive process of decoherence
or a progressive decoherence. The very sharp peaks in $D(t)$ curves \ may
originate from the reversibility of the Schroedinger equation for few body
system and we called them quantum jumps\cite{sun1,sun2}

It is very interesting to observe that the component Hamiltonian $H_{1}$ and 
$H_{0}$ are of Hermitian quadratic form of creation and annihilation
operators. Mathematically, they are the same as that to produce the
degenerate parametric amplifier in nonlinear quantum optics with classical
pump \cite{scully}. This fact tells us that the component Hamiltonian $H_{0} 
$ and $H_{1}$ can create different squeezing of the LC mode. Namely, $H_{0}$
and $H_{1}$ may drive the LC oscillation mode from the same coherent state $%
|\alpha \rangle $ to evolve into two different squeezed states\cite{yuen}.
With this mathematical consideration, we can evaluate the time evolution of
the total system and obtain the squeezing wave function at time $t$ 
\begin{equation}
|s_{k}(t)\rangle =\exp [i(-1)^{k}\frac{\omega _{a}}{2}t]|\alpha ,\mu
_{k}(t),\nu _{k}(t)\rangle _{A_{k}}
\end{equation}%
of the LC circuit for $\mu _{k}(t)[\nu _{k}(t)]=\frac{1}{2}(\sqrt{N_{k}}+[-]%
\frac{1}{\sqrt{N_{k}}})\exp (+[-]i\Omega t),$ $N_{0}=\frac{1}{N_{1}}=\sqrt{%
\frac{\omega \Delta }{\omega \Delta +4g^{2}}}$ and $\Omega =\sqrt{\omega
^{2}+4g^{2}\omega /\Delta }.$Here, the squeezed coherent state $|\alpha ,\mu
_{k}(t),\nu _{k}(t)\rangle _{A_{k}}$ is defined as in ref.\cite{yuen} for a
new set of boson operators $A_{k}=\mu _{k}(t)a-\nu _{k}(t)a^{\dag }$ (for $%
k=0,1$).

The above calculation demonstrates that the off-resonance interaction
between the LC circuit oscillator mode and the different charge qubit will
result in a dynamic squeezing split of the quasi-classical state $|\alpha
\rangle $ of LC circuit. The two split components with different squeezing
are represented by different squeezing states. Correspondingly, the
decoherence factor characterizing quantum decoherence is%
\begin{equation}
D(t)=G(t)\exp (\frac{-8g^{4}\sin ^{2}\Omega t}{\Delta ^{2}\Omega
^{2}+8g^{4}\sin ^{2}\Omega t}|\alpha |^{2})
\end{equation}%
where $G(t)=\frac{\Delta \Omega }{\sqrt{\Delta ^{2}\Omega ^{2}+8g^{4}\sin
^{2}\Omega t}}.$Considering $\frac{g}{\Delta }\ll 1$ and $\frac{g}{\Omega }%
\sim \frac{g}{\omega }\ll 1$, we can simplify the above result as 
\begin{equation}
D(t)=\exp (\frac{-8g^{4}\sin ^{2}\Omega t}{\Delta ^{2}\Omega ^{2}}|\alpha
|^{2})
\end{equation}%
The reversible decoherence phenomenon with quantum jumps illustrated in
Fig.2 is quite typical. It was found even theoretically in reference\cite%
{sun1,sun3} in 1993, and the possibility of implementing its observation in
cavity QED experiment was also pointed out in ref.\cite{sun2}. In 1997 it
was also independently discussed \cite{brune} with another cavity QED setup,
whose Hamiltonian is mathematically similar to that in our present
investigation. As understood usually \cite{zhang}, the quantum decoherence
reflects a complementarity effect since the LC mode plays a role of carrying
away information about the phase of Josephson Junction qubit and the phase
uncertainty appears when enough information of qubit is determined by the LC
mode in a very classical state. The more exact information about the qubit
phase we obtain the stronger influence will the LC mode exert on the qubit.
This revival of decoherence or quantum jump substantially results from the
fact that the reservoir is only of a single mode,\ and its profound origin
is the reversibility of the time -evolution for the system of few degrees of
freedom is reversible since governed by Schr\"{o}dinger equation. 
\begin{figure}[tbp]
\caption{{\protect\small {The time-dependence of decoherence factor with
different $|\protect\alpha |=5$(dot line),$|\protect\alpha |=10$(dash line),$%
|\protect\alpha |=30$(solid line). The larger $|\protect\alpha |$ means the
more exact "detection " about this qubit or the one-mode reservoir is more
classical. It leads to an evident vanishing of coherence. }}}
\end{figure}

In comparison with the case of atomic cavity QED, the advantage using
Josephson charge qubit to test one-bit reservoir induced decoherence is due
to the macroscopically quantum effect of superconductive system and the
well-controlled nature of coupling to one-bit engineered reservoir. A direct
way to observe the quantum jump effect of engineered quantum decoherence is
to detect the current through the probe junction as in the schematics of
Copper pair box in Fig.3. The box electrode is connected to an inductor L
via the two junctions of SQUID. When the charge qubit is in the high level
state $|1\rangle _{c}$, there are two electrons passing the probe junction.
In fact, under a proper bias condition, the state decays into $|0\rangle
_{c} $ via two single-electron tunnelling through the probe junction.

\begin{figure}[tbp]
\caption{Schematic of Cooper pair box with probe junction.The additional
voltage biased probe electrode of voltage $V_b$ is attached to the box
through a highly-resistive tunnel junction H for the detection of charge
qubit state .}
\end{figure}

As usual, it is difficult to observe the two electrons via a single trial,
but one can see an average effect of this tunnelling process. The current is
proportional to the charging rate of the occupation probability $%
P_{c}(t)=Tr(\rho |1\rangle _{cc}\langle 1|)$ of Cooper pair in $|1\rangle
_{c}$. The corresponding current $I(t)=\frac{\partial }{\partial t}%
(-2eP_{c}(t))$ is explicitly expressed for $c_{0}=c_{1}=\frac{1}{\sqrt{2}}$%
as 
\begin{equation}
I(t)\simeq e\sin \theta D(t)[\omega _{a}\sin \omega _{a}t+\frac{%
8g^{4}|\alpha |^{2}}{\Delta ^{2}\Omega ^{2}}\sin 2\Omega t\cos \omega _{a}t]
\end{equation}
\begin{figure}[tbp]
\caption{The Rabi oscillation of the charge current. (a)with coupling to the
LC oscillator($|\protect\alpha |=30$).(b)without coupling to the LC
oscillator}
\end{figure}
\noindent where we have considered the approximation $\omega _{a},\omega \gg
g$. In Fig.4, we compare this result with the case without coupling to LC
circuit. It can be seen that the current oscillates sinusoidally in both
cases, but the coupling to external reservoir adds the periodical amplitude
modulation as the direct manifestation of decoherence. Experimentally, one
can use the ratio of envelop width and the fixed period to measure the
extent of decoherence quantitatively.

In principle this quantum decoherence is macroscopically observable and it
is expected to be implemented in the experiment of Josephson qubit in the
near future. It is crucial for the above arguments to initially prepare the
L-C mode in a coherent state. As usual the external sources can add the
linear forces $\propto q$ or $\phi $ . They may force the L-C mode to evolve
into a coherent state from a vacuum state. In practice, the initial state
may easily be in a thermal equilibrium at finite temperature, but this state
is described by a diagonal density matrix in the coherent-state
representation ("Q-representation"). Thus, the quantum jump phenomenon
predicted above can still be observed and the higher temperature can enhance
the quantum jump. For the cavity QED case we have shown this enhancement
effects by straightforward calculations\cite{sun2}. The same calculations
can be done here for the charge qubit.

A difficulty to realize this setup is to fabricate a nanometer-scale
inductor with tunable inductance L. Another difficulty lies in the quantum
dissipation of the inductor causing the energy relaxation and the additional
decoherence simultaneously. The mechanism of this dissipation is due to the
coupling of the inductor to the vacuum electromagnetic field. For the
practical purpose, we shall include this dissipation effect in our future
argument.

We finally remark that the relevant quantum measurement problem of Josephson
Junction qubit has been considered theoretically by Averin\cite{averin}. He
extends the concept of quantum non-demolition (QND) measurement to coherent
Rabi oscillation of JJ qubit. The advantage of such QND measurement is that
the observation of oscillation spectrum, in principle, avoids the detector
induced decoherence. This suggested that a scheme combining flux and charge
qubit may be used in our setup to detect the engineered quantum decoherence
without "additional quantum decoherence".

{\it This work is supported by the NSF of China and the knowledged
Innovation Program (KIP) of the Chinese Academy of Science. It is also
founded by the National Fundamental Research Program of China with No
001GB309310. We also sincerely thank D.L.Zhou for the useful discussions
with him.}



\begin{references}
\bibitem[a]{email} E-mail : suncp@itp.ac.cn; web site: http:// www.
itp.ac.cn/\symbol{126}suncp

\bibitem{wheeler-zurek} J. A. Wheeler and Z.H. Zurek, Quantum Theory of
Measurement. (Princeton University Press, NJ, 1983).

\bibitem{caldeira-legget} A. O. Caldeira and A. J. Legget, Physica A, 121,
587(1983); E. Joos and H. D. Zeh, Z. Phys. B {\bf 59}, 223(1985).

\bibitem{zurek1} W. H. Zurek, Physics Today, ;Phys. Rev. D 24, 1516 (1981);
R. Onnes, Rev. Mod. Phys, {\bf 64}, 339 (1992).

\bibitem{sun1} C. P. Sun, Phys. Rev. A {\bf 48}, 878 (1993). C. P. Sun {\it %
et.al}, Fortschr. Phys. {\bf 43}, 585 (1995).

\bibitem{brune} M. Brune,{\it et.al}, Phys. Rev. Lett {\bf 77}, 4887 (1996);
J. M. Raimond, M. Brune and S Haroche, Phys. Rev. Lett {\bf 79}, 1964 (1997).

\bibitem{Zhu} H.B.Zhu,C.P.Sun, Chinese Science (A) 2000.10 30(10) 928-933;
Progresse in Chinese Science, 2000.60 10(8) 698-703 0

\bibitem{maytt} C. J. Myatt, {\it et.al}, Nature, {\bf 403}, 269 (2000).

\bibitem{shnirman} A. Shnirman, G. Sch\"{o}n, Z. Hermon, Phys. Rev. Lett, 
{\bf 79}, 2371 (1997);Y. Makhlin, G. Sch\"{o}n, A. Shnirman, Nature, {\bf 398%
}, 305 (1999).

\bibitem{nakamura} Y. Nakamura, C. D. Chen, J. S. Tsai, Phys. Rev. Lett {\bf %
79}, 2238 (1997); Y. Nakamura, Y.A. Pushkin, J. S. Tsai, Nature, {\bf 398},
786 (1999).

\bibitem{mooij} J. E. Mooij {\it et.al}, Science, {\bf 285}, 1036 (1999); C.
H. Van der Wal, science {\bf 290}, 773 (2000); J. R. Friedman, Nature, {\bf %
406}, 43 (2000).

\bibitem{vion} D. Vion, {\it et.al}, Science. {\bf 296}, 1886 (2002).

\bibitem{yu} Y. Yu, S. Han, X. Chu, S. I. Chu, Z. Wang, Science, {\bf 296},
889 (2002); S. Yan, Y. YU, X. Chu, S. I. Chu. Z.wang, Science, {\bf 293},
1457 (2001).

\bibitem{mooij1} Caspar H. van der Wal, F.K. Wilhelm, C.J.P.M. Harmans, J.E.
Mooij,LANL e-print ,cond-mat/0211664.

\bibitem{maklin} Y. Makhlin, G. Sch\"{o}n, A. Shnirman, Rev. Mod. Phys {\bf %
73}, 357 (2001).

\bibitem{sun2} C. P. Sun, {\it et.al.}, Quantum Semiclassic Opt {\bf 9}, 119
(1997).

\bibitem{zhang} P. Zhang, X. F. Liu, C. P. Sun, Phys. Rev. A {\bf 66},
042104 (2002).

\bibitem{scully} M. O. Scully, M. S. Zubairy, Quantum Optics ( Cambridge
University Press , England, 1997); D. F. Walls, G. J. Milburn, Quantum
Optics ( Springer-Verlag, Berlin, 1994).

\bibitem{yuen} H. P. Yuen, Phys. Rev. A {\bf 13}, 2226 (1976); D. F. Walls,
Nature, {\bf 324}, 210 (1981) ; C. M. Caves, Phys. Rev. D {\bf 23}, 1693
(1981).

\bibitem{sun3} C. P. Sun, in Quantum Coherence and Decoherence. ed. by K.
Fujikawa and Y. A. Ono, (Amsterdam; Elsevier Science Press, 1996).

\bibitem{averin} D. V. Averin, e-print LANL Cond/0202082 VI, 2002; fortsch
Physik, {\bf 48}, 1055 (2002).
\end{references}
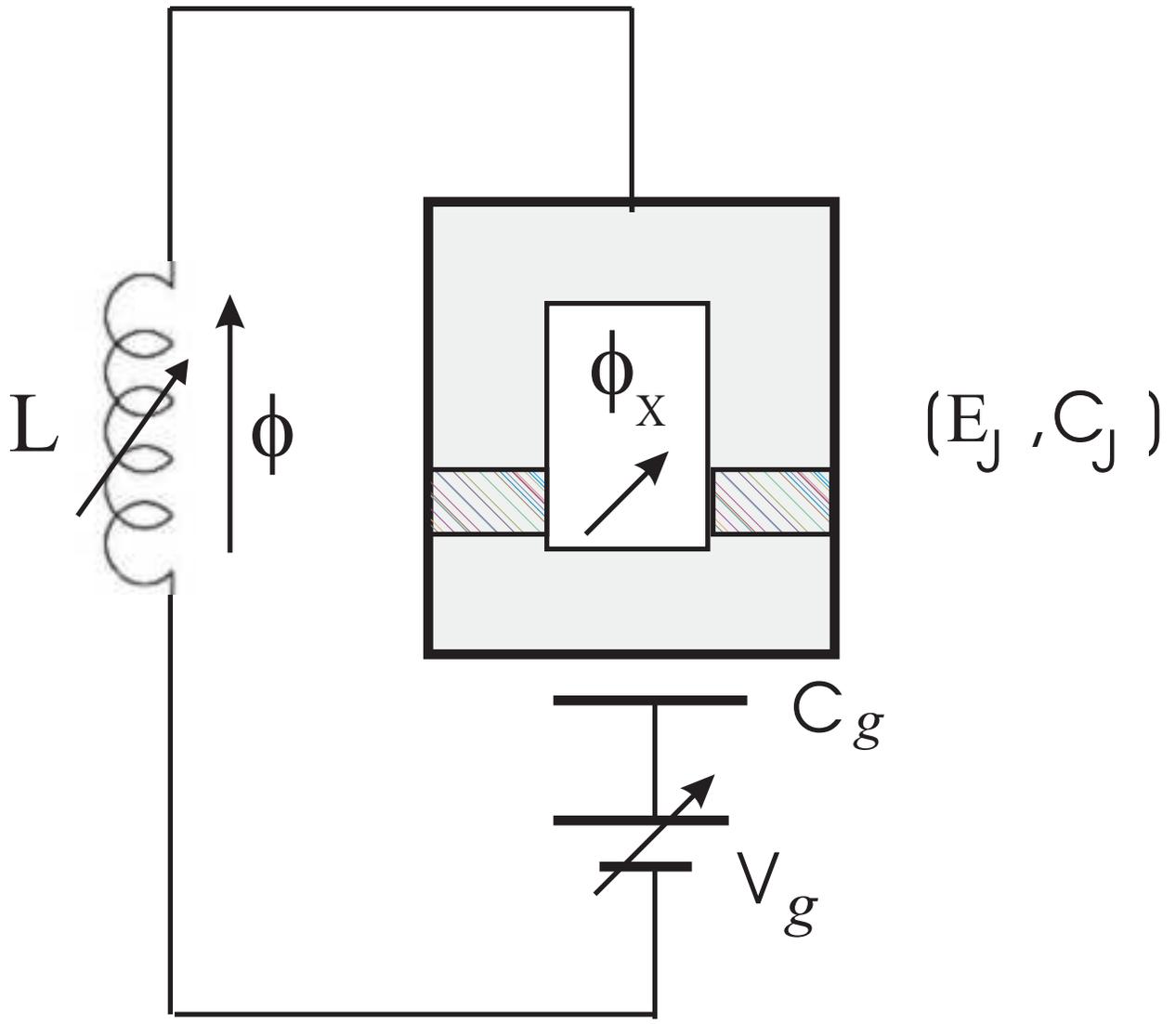
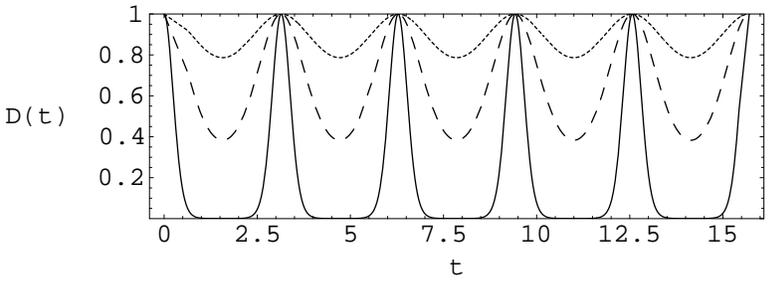
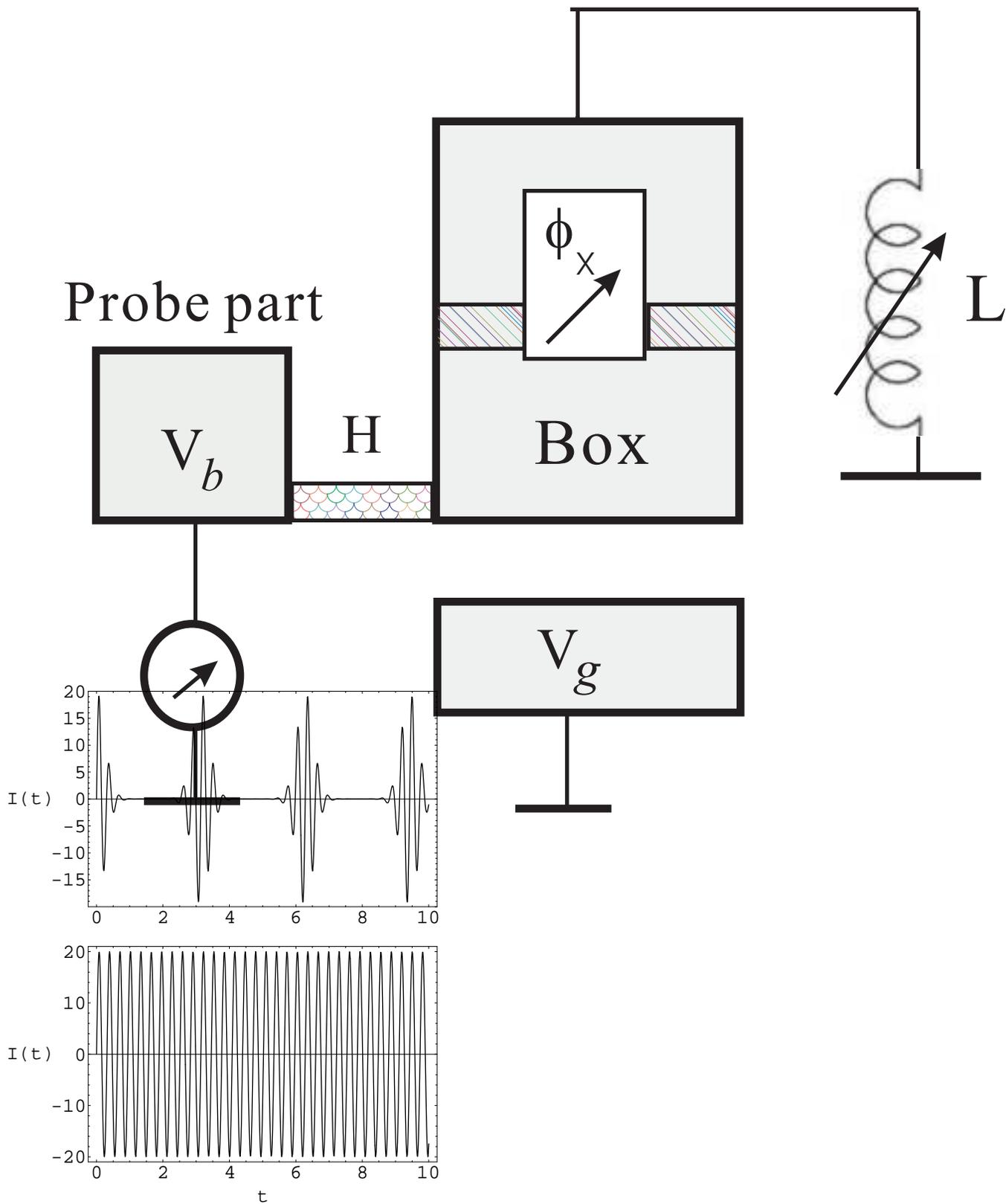
\end{document}